\documentstyle[12pt,aasms4]{article}

\lefthead{Wang} \righthead{On the Possible Nature of the Unidentified
 Source RX J0720.4-3125 and its Implications}

\begin{document}

\title{Evidence for magnetic field decay in RX J0720.4-3125}

\author{John C. L. Wang} \affil{Department of Astronomy, University of
Maryland, College Park, MD 20742-2421}


\begin{abstract}
The unidentified X-ray source RX J0720.4-3125 is a candidate
 isolated neutron star showing evidence for {\it pulsed\/} emission 
with an 8.39 s period and a spectrum consistent with a blackbody
 at $kT=80$ eV (Haberl {\it et al.\/} 1996, 1997).
 We show that this source is most likely an old isolated neutron star
 accreting from surrounding media. We then argue
 that unless it was born with a long spin period
 ($P_i\gtrsim 0.5$ seconds) {\it and\/} weak field ($B_i\lesssim 10^{10}\,G$),
 the magnetic field on this star {\it must\/} have decayed.
 With $B_i\sim 10^{12}\,G$, we find decay timescales
 $\gtrsim 10^7$ yrs for power law decay or $\gtrsim 10^8$ yrs
 for exponential decay. A measured period derivative 
 ${\dot P}\lesssim 10^{-16}$ s$\,$s$^{-1}$ would
 be consistent with an old accreting isolated neutron star.
  Both power law and exponential decay models can give 
  a ${\dot P}\sim 10^{-16}$ s$\,$s$^{-1}$, though a ${\dot P}$ substantially
  less than this would be indicative of exponential field decay.

\end{abstract}

{\it Subject Headings:} pulsars: general --- stars: individual (RX J0720.4-3125) ---

   stars: magnetic fields --- stars: neutron --- X-rays: stars

\section{Introduction}
\label{intro}

There are expected to be $\sim 10^8$-$10^9$ isolated neutron 
stars created in the Galaxy since its formation
(e.g., Shapiro \& Teukolsky 1983).
  The vast majority of these stars are expected to have spun down 
  from their initial short spin
 periods and to have long ceased being active pulsars.
 If, however, they can accrete from surrounding media, 
 they can become visible as sources of soft 
 quasi-thermal X-rays (Shvartsman 1970; Ostriker, Rees, Silk 1970; Treves \& Colpi
 1991; Blaes \& Madau 1993; Madau \& Blaes 1994).  Compared to the X-ray 
 luminosity, these objects are expected to be very weak in the optical and 
 infrared (i.e., typically $L_{X}\sim L_{tot}\sim 10^{30}$ erg/s and
 $L_{opt,IR}/L_X\ll1$) regardless of whether this emission originates from the 
 stellar surface or from the surrounding photoionized nebula (cf. Blaes \& Madau 
 1993; Blaes et al. 1995). 
 A detection of such old neutron stars would significantly advance
 our understanding of their spin and magnetic field evolution.
 The magnetic field evolution of isolated neutron stars
 is a major unresolved issue in compact object astrophysics.
 Theoretical studies lead, on the one hand, to 
 exponential (Ostriker \& Gunn 1969) or power law (Sang \& 
 Chanmugam 1987; Goldreich \& Reisenegger 1992; Urpin et al. 1994) 
 forms of field decay, to little or
 no decay within the age of the universe 
  (Romani 1990; Srinivasan et al. 1990; Goldreich \& Reisenegger 1992) 
 on the other.
 Statistical studies based upon the observed isolated radio pulsars
 give equally equivocal results (Lyne, Anderson, Salter 1982;
 Narayan \& Ostriker 1990; Sang \& Chanmugam 1990; Bhattacharya et al.
 1992; Verbunt 1994) owing in large part to the difficulty in treating
 strong selection effects (e.g., Lamb 1992).

 In this paper, we first argue that the soft X-ray source RX J0720.4-3125 
 is an isolated neutron star accreting from its surroundings. Unlike the 
 two other candidate old neutron stars (cf. Stocke et al. 1995; 
 Walter et al. 1996), this source shows evidence for an 8.39 s rotation 
 period --- longer than in any known radio or $\gamma$-ray pulsar. From this,
 we argue that the magnetic field on this star {\it must\/} decay on timescales 
 $\gtrsim 10^7$ yrs if it was born spinning rapidly. A measurement of the period 
 derivative ${\dot P}$ would help test this model.

\medskip

\section{Observations}
\label{obs}

\medskip

  The source RX J0720.4-3125 is an unidentified soft X-ray source seen
  by the Einstein IPC (Image Proportional Counter), EXOSAT LE (Low 
  Energy Detector), and, most recently, by the ROSAT PSPC (Position
  Sensitive Proportional Counter) and HRI (High Resolution Imager). 
  The data on this source is as follows (Haberl et al. 1996, 1997):
  The ROSAT PSPC count rate (0.1 -- 2.4 keV) is 1.6 cts/s, 
  and based upon earlier detections
  by Einstein and EXOSAT, is steady (no more than $\pm10$\% variation) over many 
  years. The X-ray spectrum is best fit by a black body of $kT=80$ eV 
  with a hydrogen absorption column density of 
  $N_H=1.3\times10^{20}\,\,$ cm$^{-2}$. With this spectrum, the count
  rate corresponds to an unabsorbed photon energy flux of 
  $F_\nu(0.1-2.4\,\,{\rm keV}) \approx 1.7\times 10^{-11}$ erg/cm$^2$/s 
  (K. Arnaud, priv. comm.).
  The source luminosity is then

\begin{equation}
L_X \equiv L(0.1-2.4\,\,{\rm keV}) = 1.9\times10^{31}d_{100}^2\ \ {\rm erg/s},
\label{lumx}
\end{equation}

 \noindent where $d=100d_{100}$ pc is the distance to the source.

  In all pointed ROSAT observations, there is a {\it periodic\/} 
  modulation in the X-ray flux with an 8.39 s period
  (Haberl et al. 1996, 1997). We interpret this as the rotation
  period of the source. Since the ROSAT observations are separated
  by $\sim 3$ years, this indicates that the pulsed emission is steady.
   Two pointed ROSAT HRI observations (Haberl et al. 1996, 1997)
   secured the position of the source to be (J2000) 
   $\alpha=$ 7h, 20m, 24.90s; $\delta=$-31$^\circ\,$ 25$^\prime\,$
    51.3$^{\prime\prime}$
   (with $\pm3^{\prime\prime}$ uncertainty).
   The corresponding Galactic coordinates are
   $l=244^\circ,\,b=-8^\circ$.
    Optical observations at the South African Astronomical Observatory
    failed to detect an optical counterpart down to a limiting 
    magnitude of $V\sim 20.7$, thereby placing
    a lower limit on the X-ray to optical flux ratio of $\sim 500$
    (Haberl et al. 1997).

  The observational evidence points consistently
  to an isolated neutron star as the source.

\section{The distance to RX J0720.4-3125}
\label{dist}

\medskip

  We estimate the distance to the source from the (low) 
  hydrogen column density ($N_H=1.3\times10^{20}$ cm$^{-2}$). 
  For the first $\sim 150$ pc, the line of sight to this source
  cuts through the Local Bubble where the mean hydrogen density 
  is $n_H\sim 0.05$ cm$^{-3}$ (Welsh et al. 1994, their Figure 3). 
  Beyond the Local Bubble, the mean hydrogen density increases 
  substantially to $n_H\sim 0.5$ cm$^{-3}$ near the Galactic plane
  (Dickey and Lockman 1990). 
  Taking an empty Local Bubble gives a rough upper 
  bound to the distance of about 250 pc.

   Given the very nonuniform matter distribution in the local interstellar
   medium (e.g., Welsh et al. 1994), the actual distance could be much less 
   than 250 pc. For instance, if the source intercepts diffuse cirrus, its 
   distance could be closer to $\sim 100$ pc (cf. Wang and Yu 1995).
  A very conservative but strict upper bound on the distance is set 
  by requiring the hot spot area to be much less than the star's
  surface area for pulsations to be observed. This implies 
  $d_{100}\ll 5.3R_{6}$, where $R=10^6R_{6}$ cm is the stellar radius.
   For definiteness, we adopt 100 pc throughout this work
   as the source distance and scale our results to this value.
  Our conclusions regarding magnetic field decay are not sensitive 
  to the distance estimate.

\section{Arguments against a young neutron star/pulsar}
\label{yns}

\medskip

  An active pulsar's spin-down power is
  (e.g., Shapiro \& Teukolsky 1983)
${\dot E_R} = {{8\pi^4B^2R^6}\over{3c^3P^4}}$,
  where $B$ is the dipole magnetic field strength at the 
  polar cap. (We took $\sin\alpha=1$, where $\alpha$ is the angle 
  between the rotation axis and magnetic dipole moment
  [cf. Goldreich \& Julian 1969; Verbunt 1994].)

   For young radio pulsars (age$\sim 10^4$--$10^6$ yrs), ${\dot E_R}\gg L_X$ 
   (cf. eqn [\ref{lumx}]; e.g., \"Ogelman \& Finley 1993), yielding
   $B_{12} \gg 140d_{100}R_6^{-3}P_{8.39}^2$, where $P=8.39P_{8.39}$ s is 
   the current observed pulsar period and $B=10^{12}\,B_{12}\,G$. 
  This qualifies the source as a ``magnetar'' (Duncan \& Thompson 1992). 
  The spin-down rate is ${\dot P}\Big\vert_{now} (s\,s^{-1})= 
  2.4\times10^{-16}\,{{B_{12}^2R_6^6}\over{I_{45}P}} \gg
  6\times10^{-13}\,d_{100}^2 P_{8.39}^3I_{45}^{-1}$,
  where $I=10^{45}I_{45}$ g-cm$^2$ is the star's moment of inertia.
    The consequent young spin down age --- 
    $\tau_{sp} = {{P }\over{2{\dot P} }} \ll  2.2\times10^5\,d_{100}^{-2}
    P_{8.39}^{-2}I_{45}$ yrs --- and close proximity ($\sim 100$ pc)
    argues against a young neutron star/pulsar origin.
     This object lies well above the (extrapolated)
     observed radio pulsar death line 
     ($B_{12} > 15P^2_{8.39}$; e.g., Chanmugam 1992).
   Adopting $L_{radio}$ (mJy-kpc$^2) = 4\times10^6\, {\dot P}^{1/3}/P$ 
   (cf. Narayan \& Ostriker 1990) yields a radio luminosity 
   at 400 MHz of $L_{400} \sim 4300\,d_{100}^{-{4/3}}I_{45}^{-{1/3}}\ \ {\rm mJy}$.
  Unless our line-of-sight falls outside its radio beam
  (J. Cordes, priv. comm.), then even given the large spread 
  (up to a factor of 100) in the actual radio luminosities about the 
  best-fit $L_{radio}(P, {\dot P})$ (e.g., Lamb 1992), such a 
  bright and {\it persistent\/} radio pulsar should have already been 
  detected (cf. Taylor, Manchester, Lyne 1993).
   The absence of radio emission further argues against a 
   young neutron star/pulsar origin.

  If the source is a Geminga-type ($\gamma$-ray loud, radio-quiet) pulsar, 
  then ${\dot E_R}\sim L_\gamma\gg L_X$.
  Adopting $L_\gamma/L_X \sim 10^3$ as for Geminga (Halpern \& Holt 1992;
  Swanenburg et al. 1981; Thompson et al. 1977) yields $\tau_{sp} \sim  200$ yrs,
  which argues against a Geminga-type pulsar.
   This is consistent with lack of detection by EGRET 
   instruments aboard the Compton Gamma-Ray Observatory
   (Haberl et al. 1997).

\section{Accreting old neutron star and field decay}
\label{ons}

\medskip

If the source is an isolated neutron star accreting from the surrounding
medium, its mass accretion rate will be governed by the rate at which 
material is captured within the star's gravitational radius 
  $r_g = {{2GM}\over{V^2}} = 9.3\times10^{13}\,M_{1.4}V_{20}^{-2}\, {\rm cm}$,
  and is given by (cf. Bondi 1952)
 ${\dot M}_{11} = 1.3M_{1.4}^2n_HV_{20}^{-3}f_{acc}$,
 where ${\dot M}=10^{11}{\dot M}_{11}$ g/s, 
 $M=1.4M_{1.4}M_\odot$ is the stellar mass, 
 $n_H$ is the hydrogen number density of the medium 
  (assumed to have solar abundance), 
  $V=(v^2+c_s^2)^{1/2}=20V_{20}$ km/s with $v$ being the star's
 speed relative to the ambient medium, $c_s$ being the sound speed
 in the ambient medium, and $f_{acc}$ is a factor
 that accounts for the microphysics of the accretion process. 
  If the 
 accretion is adiabatic, 
   $f_{acc}\sim 1$ (Bondi-Hoyle accretion). 
 If preheating of the 
 incident flow is important, $f_{acc}$ could be much less than 
 unity (Shvartsman 1970; Ostriker et al. 1976; Blaes, et al. 1995; 
 Wang \& Sutherland 1997).
  Combining eqn (\ref{lumx}) for $L_{tot}$ and $L_{tot}=GM{\dot M}/R$
  gives $V_{20} = 1.1 f_{acc}^{1/3}n_H^{1/3}d_{100}^{-{2/3}}$.
 Taking $c_s\sim 10$ km/s then gives $v \lesssim 2c_s$,
 that is, a slowly moving neutron star.
   In this case,
  accretion occurs in a quasi-spherical 
  manner near the star (i.e., for $r_g\gg r\gg $ Alfv\'en radius [see below];
  cf. Hunt 1971).
      The condition $v>0$ also gives another distance constraint; 
      $d_{100}<3.3f_{acc}^{1/2}n_H^{1/2}$, with the upper limit
      corresponding to $v=0$.

 For material to reach the surface of the rotating magnetized star,
 the star must have spun down sufficiently so that accreting material
 can overcome the centrifugal barrier (cf. Illarionov \& Sunyaev 1975).
   The neutron star's spin evolution is divided into three phases 
   (see, e.g., Blaes \& Madau 1993; Lipunov, Postnov, \& Prokhorov 1997).
  In the first (dipole) phase,
  the star is an active pulsar and spins down by magnetic 
  dipole radiation;
 $-{\dot \Omega} = -{\dot\Omega}_{dip} = {{B^2R^6\Omega^3}\over{6Ic^3 }}
               = 6.2\times10^{-18}R_6^6I_{45}^{-1}
                  B_{12}(t)^2\Omega(t)^3\ \ 
                 ({\rm s}^{-2})$,
  where $\Omega=2\pi/P$ is the star's angular velocity.
 This phase ends when the ram pressure of the ambient material
 ($\sim\rho V^2$) overcomes the pulsar wind pressure 
 ($\sim {\dot E}_R/(c4\pi r^2)$) at $\sim r_g$
 so that matter can now enter the star's magnetosphere.
 This happens when the star's period 
 $P > P_0 \equiv 4.4M_{1.4}^{-{1/2}}R_6^{3/2}v_{20}^{1/2}n_H^{-{1/4}}
                   B_{12}^{1/2},\ {\rm s}$.
  In this second (propeller) phase, material enters the 
  corotating magnetosphere and is stopped at $\sim r_A$, the Alfv\'enic 
  radius, where the energy density in the accretion flow balances the 
  local magnetic pressure. This radius is given by
 $r_A = 1.5\times10^{10}f_{acc}^{-{2/7}}n_H^{-{2/7}}V_{20}^{6/7}
                  M_{1.4}^{-{5/7}}\mu_{30}^{4/7}\ {\rm cm}$,
 where $\mu = BR^3/2 = 10^{30}\mu_{30}\,$ G-cm$^3$ is the neutron
 star's magnetic moment.

  Further penetration cannot occur
  owing to the centrifugal barrier, that is, $r_A>r_{co}$
  (cf. Illarionov \& Sunyaev 1975), 
  where
 $r_{co} = \left({{GM}\over{\Omega^2}}\right)^{1/3} =
           6.9\times10^{8}M_{1.4}^{{1/3}}P_{8.39}^{2/3}\ \,{\rm cm}$
   is the corotation radius.
  Since the accretion is quasi-spherical, material falls in initially
  with negative total energy (material is bound) and roughly zero 
  angular momentum.

  Spin down in this phase occurs via propeller and
  magnetic dipole spin-down;
   ${\dot \Omega} = {\dot\Omega}_{prop} + {\dot\Omega}_{dip}$.
  If the infalling
  material cools efficiently and attaches itself to field lines 
  at around $r_A$, the star expels the material once it
  spins up the material to the local escape velocity at $\sim r_A$.
  By angular momentum conservation,
   $-{\dot \Omega}_{prop}^{l} \sim {{{\dot M}(2GMr_A)^{1/2}}\over{I}}
            = 2.5\times10^{-16}M_{1.4}^{15/7}R_6^{6/7}I_{45}^{-1}
                 f_{acc}^{6/7}n_H^{6/7}V_{20}^{-{18/7}}B_{12}(t)^{2/7}\ ({\rm s}^{-2})$.
    If the material does not cool and/or thread
    the field lines efficiently, the hot gas will remain bound 
    in a ``cocoon'' at $\sim r_A$.
    As the underlying magnetosphere rotates supersonically shearing 
    through this ``cocoon,''
    the consequent shock heating expels material from the star
     (Illarionov \& Sunyaev 1975).
        By energy conservation,
   $-{\dot \Omega}_{prop}^{e} \sim { {GM{\dot M} } \over {I\Omega r_A} }
                 = {1\over{\sqrt{2}}}\left({ {r_{co} }\over{r_A } }\right)^{3/2}
                      {\dot \Omega}_{prop}^{l} 
                 \sim 10^{-2} {\dot \Omega}_{prop}^{l}$,
    where the final expression is typical for a star just entering the propeller
    phase. In this limit, only a fraction of the energy given to escaping material
    goes into azimuthal motion, 
     so $-{\dot\Omega}_{prop}^e\ll-{\dot\Omega}_{prop}^l$.
    We therefore take
    ${\dot\Omega}_{prop}=f_p{\dot\Omega}_{prop}^{l}$,
     where $0.01\lesssim f_p\lesssim 1$
     is taken as a freely adjustable parameter.
     The exact value of $f_p$ requires time-dependent numerical
     simulations (e.g., Wang \& Robertson 1985).
    If $f_p\sim 1$,
    ${\dot\Omega}_{prop}\gg{\dot\Omega}_{dip}$ throughout 
    this phase.

  Propeller action continues until $r_A<r_{co}$, when
  the centrifugal barrier is removed and polar cap accretion ensues
  (e.g., Lamb et al. 1973; Davidson \& Ostriker 1973; Arons \& Lea 1980).
  This occurs when
  $P > P_a \equiv 470M_{1.4}^{-{11/7}}R_6^{18/7}f_{acc}^{-{3/7}}
                    n_H^{-{3/7}}V_{20}^{9/7}B_{12}^{6/7} \ {\rm s}$.
   In quasi-spherical accretion, no net {\it torque\/} is exerted on the star 
   (due to accretion). 
     The mass loading of the field lines, however,
     now becomes important in spinning down the star by
     increasing the moment of inertia of the star + corotating
     magnetosphere system
       {\footnote{I thank Eve Ostriker for pointing out the potential
                  importance of this effect.}}
   (see, e.g., Mestel 1990 for a discussion of this effect in normal stars).
    For $r_A\gg R$, angular momentum conservation (${{d}\over{dt}}(I\Omega)=0$)
    gives $-{\dot \Omega}_{brk} = {{\dot I }\over{I }}\,\Omega
                        \sim {{{\dot M}r_A^2 }\over{MR^2 }}\,\Omega
                      = 4.5\times 10^{-15}\,
                         M_{1.4}^{-{3/7}}R_6^{10/7}n_H^{3/7}
                         V_{20}^{-{9/7}}f_{acc}^{3/7}\,B_{12}^{8/7}\,\Omega
                            \ ({\rm s}^{-2})$.
    (In the limit of a weakly magnetized star where $r_A<R$, 
      $r_A\to R$ in the above expression and 
     $-{\dot \Omega}_{brk} \to {{{\dot M}}\over{M}}\,\Omega$.)
    The total spin-down rate in this final (accretion) phase is then 
    ${\dot \Omega} = {\dot \Omega}_{brk} + {\dot \Omega}_{dip}$,
    although, quite generally, ${\dot \Omega}_{brk} \gg {\dot \Omega}_{dip}$.

    For accretion to occur, we require $r_A<r_{co}$, that is, $P>P_a$.
     For pulsed emission, that is, polar cap accretion, we require $r_A\gg R$.
      Combining these, the present day surface field must satisfy 

\begin{equation}
    10^{-7}\,M_{1.4}^{5/4}R_6^{-{5/4}}V_{20}^{-{3/2}}n_H^{1/2}f_{acc}^{1/2} \ll
 B_{12} < B_{crit,12} \equiv 9.3\times10^{-3}M_{1.4}^{11/6}R_6^{-3}f_{acc}^{1/2}
           V_{20}^{-{3/2}}n_H^{1/2}P_{8.39}^{7/6}.
\label{bprop}
\end{equation}

  \noindent This condition rules out the possibility of measuring the field
  directly through cyclotron (emission) line observations, which require
  $B_{12}\sim 1$ (Nelson et al. 1995).

    Assume first that the star's field does not decay, that is, its
    initial field $B_i$ is the same as the current field and satisfies
    eqn (\ref{bprop}). 
      We obtain an upper bound to the period $P_0$ by using
      $B<B_{crit}$ (cf. eqn [\ref{bprop}]);

\begin{equation}
 P_0 < P_{crit} \equiv 0.42M_{1.4}^{5/12}f_{acc}^{1/4}
                 V_{20}^{-{1/4}}\ {\rm s}.
\label{pinit}
\end{equation}

 If the star was born with $P_i<P_0<P_{crit}$, it must first dipole spin
 down to $P_0$. This takes
 $t_{dip,0} = 6.3\times10^7\,P_0^2B_{12}^{-2}R_6^{-6}I_{45}\ {\rm yrs}
                       > 1.3\times10^{11}\,M_{1.4}^{-{17/6}}I_{45}
                       P_{8.39}^{-{7/6}}n_H^{-1}V_{20}^{5/2}\ {\rm yrs}$,
 where we have used $B<B_{crit}$ (cf. eqn [\ref{bprop}])
 to arrive at the inequality. The star thus spends longer than a
 Hubble time ($\sim 10^{10}$ yrs) just in the first (dipole) phase.

    If the star was born with $P_0<P_i<P_{crit}$, it goes directly to 
    the second (propeller) phase. Propeller spin down to $P_a$ takes
 $t_{prop,a} = 1.1\times10^9\,f_p^{-1}\,M_{1.4}^{-{15/7}}R_6^{-{6/7}}I_{45}
              f_{acc}^{-{6/7}}n_H^{-{6/7}}V_{20}^{18/7}B_{12}^{-{2/7}}
              \left({1\over{P_i}}-{1\over{P_a}}\right)\ {\rm yrs}
     >10^{10}\,f_p^{-1}\,M_{1.4}^{-{37/12}}I_{45}n_H^{-1}
             f_{acc}^{-{5/4}} V_{20}^{13/4}P_{8.39}^{-{1/3}}\ {\rm yrs}$,
 where we have used $B<B_{crit}$ (cf. eqn [\ref{bprop}]), 
 $P_i<P_{crit}$ (cf. eqn [\ref{pinit}]), 
 and $P_i\ll P_a$ to arrive at the inequality. The star thus spends longer than 
 a Hubble time in this phase.
    (If $f_p\ll1$, magnetic dipole spin down will be important 
     initially, but propeller spin down dominates eventually.)

  We conclude that if the star was born with 
 $P_i \lesssim P_{crit} = 0.42M_{1.4}^{5/12}f_{acc}^{1/4}V_{20}^{-{1/4}}$ s,
  then to enable sufficient spin down to allow accretion onto
  the star, it {\it must\/} have been born with a stronger field than
  at present, that is, {\it the stellar magnetic field must have
   decayed}.

  Of course, if the neutron star is born slowly rotating, i.e, with
  $P_i\gg P_{crit}$, then the star can spin down within a Hubble time
  to 8.39 s without requiring magnetic field decay.
  However, it is unclear how neutron stars can form with such long
  initial spin periods {\it and\/} weak magnetic fields (cf. eqn [\ref{bprop}]). 
  One possibility is that RX J0720.4-3125 actually evolved from a high
  mass X-ray binary system (cf. Haberl et al. 1997).
       {\footnote{It is also possible that no accretion is involved
                  and that RX J0720.4-3125 is powered instead by some internal
                  heat source 
                  (see, e.g., Thompson \& Duncan 1996).}}
   We argue here, however, that conventional isolated neutron star 
   models can account quite well for the properties of RX J0720.4-3125.

  In general, the star's spin and magnetic field history is determined 
  by $P_i$, $B_i$, the field decay law, and decay timescale $t_d$. 
  The dependence on $P_i$, however, is very weak whenever $P_i\ll P_a$, 
  which is generally believed to be the case (e.g., Narayan \& Ostriker 1990).

 In Figure 1, we illustrate sample evolutionary tracks in $B$-$P$ space for 
 two magnetic field decay laws.  For the solid curve, the stellar field is 
 assumed to decay as a power law, i.e., $B(t)=B_i/(1+t/t_d)$ (e.g., Narayan 
 \& Ostriker 1990; Sang \& Chanmugam 1987), with $t_d=3.8\times10^7$ yrs. 
 For the dot-dashed curve, exponential decay is assumed, i.e., 
 $B(t)=B_i\,\exp(-t/t_d)$ (e.g., Ostriker \& Gunn 1969), with $t_d=4\times10^8$ yrs.  
  To construct these tracks, an isolated neutron star with $M=1.4M_\odot$ 
 and $R=10$ km is assumed to be born with 
 $P_i=0.01\ {\rm s}\, <\,P_{crit}$, $B_i=10^{12}\,G$, and is assumed
 to be moving at 20 km/s through a medium with $n_H=1$ cm$^{-3}$ and $c_s=10$ km/s.
  Spin down spans the magnetic dipole, propeller, and accretion phases.
 For the propeller spin down rate, we took $f_p=1$ and for the 
 accretion rate, we took $f_{acc}=1$.
  In the power law decay model, the star enters the accretion phase at
  $5.7\times10^9$ yrs and spins down to 8.39 s in $6.2\times10^9$ yrs.
  For the exponential decay model, these numbers are
  $1.9\times10^9$ yrs and $2.1\times10^9$ yrs, respectively.

  For $B_{i,12}\sim1$, we find $t_d\gtrsim 10^7$ yrs for power law decay 
  models (cf. Urpin et al. 1994), while $t_d\gtrsim 10^8$ yrs for exponential 
  decay models.
  These results are consistent with several recent analyses of pulsar statistics 
  and field decay (e.g., Wakatsuki et al.  1992; Bhattacharya 
  et al. 1992; Lamb 1992; Urpin et al. 1994; Verbunt 1994), and with the analysis assuming power
  law decay in Narayan \& Ostriker (1990). 
    For $B_{i,12}\lesssim 0.1$, the star cannot enter
   the accretion phase within a Hubble time even for the most favorable
   case of a stationary ($v=0$) star. 
   If $f_p\ll 1$, the star will not be able to enter the accretion phase 
   {\it and\/} spin down to 8.39 s in a Hubble time unless
   $v\ll 20$ km/s (cf. Blaes \& Madau 1993).

 From ${\dot{\Omega}}_{brk}$ and $B<B_{crit}$ (cf. eqn [\ref{bprop}]),
 we obtain an upper bound to the current spin down rate;

\begin{equation}
    {{\dot P}}\Big\vert_{NOW} < 1.8\times 10^{-16}\,
                 M_{1.4}^{5/3}R_6^{-2}n_HV_{20}^{-3}f_{acc}P_{8.39}^{7/3}
                  \ \ ({\rm s}\,{\rm s}^{-1}). 
\label{pdotnowp}
\end{equation}
  
    \noindent In both models shown in Figure 1, the 8.39 s period is reached 
    shortly after the star enters the accretion phase, and 
    ${{\dot P}}\big\vert_{NOW} \approx 10^{-16}\ {\rm s}\,{\rm s}^{-1}$.

    It is evident from Figure 1 that $P$ in the exponential decay model
    asymptotes after entering the accretion phase while $P$
    for the power law decay model continues to increase monotonically.
    This is because in exponential decay models, the field, and hence $r_A$ 
    (lever arm) decreases more rapidly at late times, whereas the field in
    power law decay models persists longer.  If neutron star
    fields undergo exponential decay and the 8.39 s period in RX J0720.4-3125 is 
    the star's asymptotic period, then we expect 
     ${\dot P}\vert_{NOW}\ll 10^{-16}$ s$\,$s$^{-1}$.

    We have so far assumed that the stellar magnetic field decays indefinitely. 
    We emphasize, however, that once the stellar field has decayed sufficiently 
    to enable accretion (cf. eqn [2]), no further decay is required by the observations. 
    Thus, for instance, the decaying field may level out at late times 
    to a steady finite long-lived value (e.g., Kulkarni 1986; Romani 1990).

\section{Conclusions}
\label{conc}

\medskip

    We argue in this paper that RX J0720.4-3125 is an old isolated neutron
    star, situated at $\sim 100$ pc, that has spun down long past the 
    active pulsar stage and is now accreting matter from the interstellar medium. 
   Unless the star was born with an unusually long period 
   ($P_i \gtrsim 0.5$ s; cf. eqn [\ref{pinit}]) {\it and\/} weak 
   magnetic field ($B_i \lesssim 10^{10}\,G$; cf. eqn [\ref{bprop}]),
   such spin down is only possible if the star's field at birth was much stronger 
   than at present, that is, the star's field {\it must\/} have decayed.
      Our analysis gives long decay timescales;
      $\gtrsim 10^7$ yrs for power law decay and $\gtrsim 10^8$ yrs for 
      exponential decay, assuming $B_i\sim 10^{12}\,G$.
          
   A ${\dot P}\lesssim 10^{-16}$ s$\,$s$^{-1}$ 
   would be consistent with an old accreting isolated neutron star.
  Both power law and exponential decay models can give 
  a ${\dot P}\sim 10^{-16}$ s$\,$s$^{-1}$. 
  A ${\dot P}\ll 10^{-16}$ s$\,$s$^{-1}$, however, 
  would be indicative of exponential field decay.
   In addition, an accreting isolated neutron star should be surrounded 
   by an extended photoionized nebula (cf. Shvartsman 1971; Blaes \& Madau 1993; 
   Blaes et al. 1995).  This low surface brightness nebula should 
   be dominated by $H\alpha$ in a compact inner zone ($\sim 10^{17}$ cm; 
   e.g., Blaes et al. 1995) and be rich in mid to far 
   infrared metal forbidden lines (e.g., [NeII]12.8$\mu$, [SiII]35$\mu$) 
   in an extended outer zone ($\sim 10^{18}$ cm).  
   The results of deep spectroscopy can therefore be used as an additional 
   test of the model presented here.

\acknowledgments
We thank G\"unther Hasinger for pointing out this source to us, and we thank
Doug Hamilton, Eve Ostriker, Jim Stone, Mark Wolfire, and Sylvain Veilleux
for valuable discussions and a critical reading of the manuscript.
We also thank an anonymous referee for insightful comments and suggestions.
This work was supported in part by NASA Astrophysics Theory Program grant 
NAG5-3836.

\newpage

\begin{figure}
\figcaption{
  The spin period and magnetic field evolutionary track
  for a neutron star ($M=1.4\,M_\odot,\,R=10$ km) 
  born with period $P_i=0.01$ s and
  surface dipole field strength $B_i=10^{12}\,G$. 
  The star is assumed to be moving at $v=20$ km/s through a medium 
  (solar abundance) with density $n_H=1$ cm$^{-3}$ and $c_s=10$ km/s.
    The star's field undergoes (solid curve) power law decay 
    ($B(t)=B_i/(1+t/t_d)$) with $t_d=3.8\times10^7$ yrs, or
    (dot-dashed curve) exponential decay ($B(t)=B_i\exp(-t/t_d)$)
    with $t_d=4\times10^8$ yrs.
    Dashed line gives the observed radio pulsar death line ($B_{12}=0.2P^2$).
  For power law decay, the star drops below the death line 
  at $t\approx 3.5\times10^8$ yrs after its birth when $P=0.73$ s.
  Propeller spindown begins at $t\approx 1.3\times10^9$ yrs when
  $P=P_0=0.76$ s. The star enters the accretion phase at
  $t\approx 5.7\times10^9$ yrs when $P=6.96$ s.
    Spin down to the observed 8.39 s period occurs after $6.2\times10^9$ yrs.
   For exponential decay, the corresponding times and periods are
   ($3.1\times10^8$ yrs, $1.56$ s), ($7.5\times10^8$ yrs, 1.74 s),
   ($1.9\times10^9$ yrs, 7.86 s), and ($2.1\times10^9$ yrs, 8.39 s), respectively.
  Each of these events is marked with an open circle on the curve for
  the power law decay model (solid) and an open triangle on the curve for the 
  exponential decay model (dot-dash).
\label{spin}}
\end{figure}

\begin{figure}
\epsscale{1}
\plotone{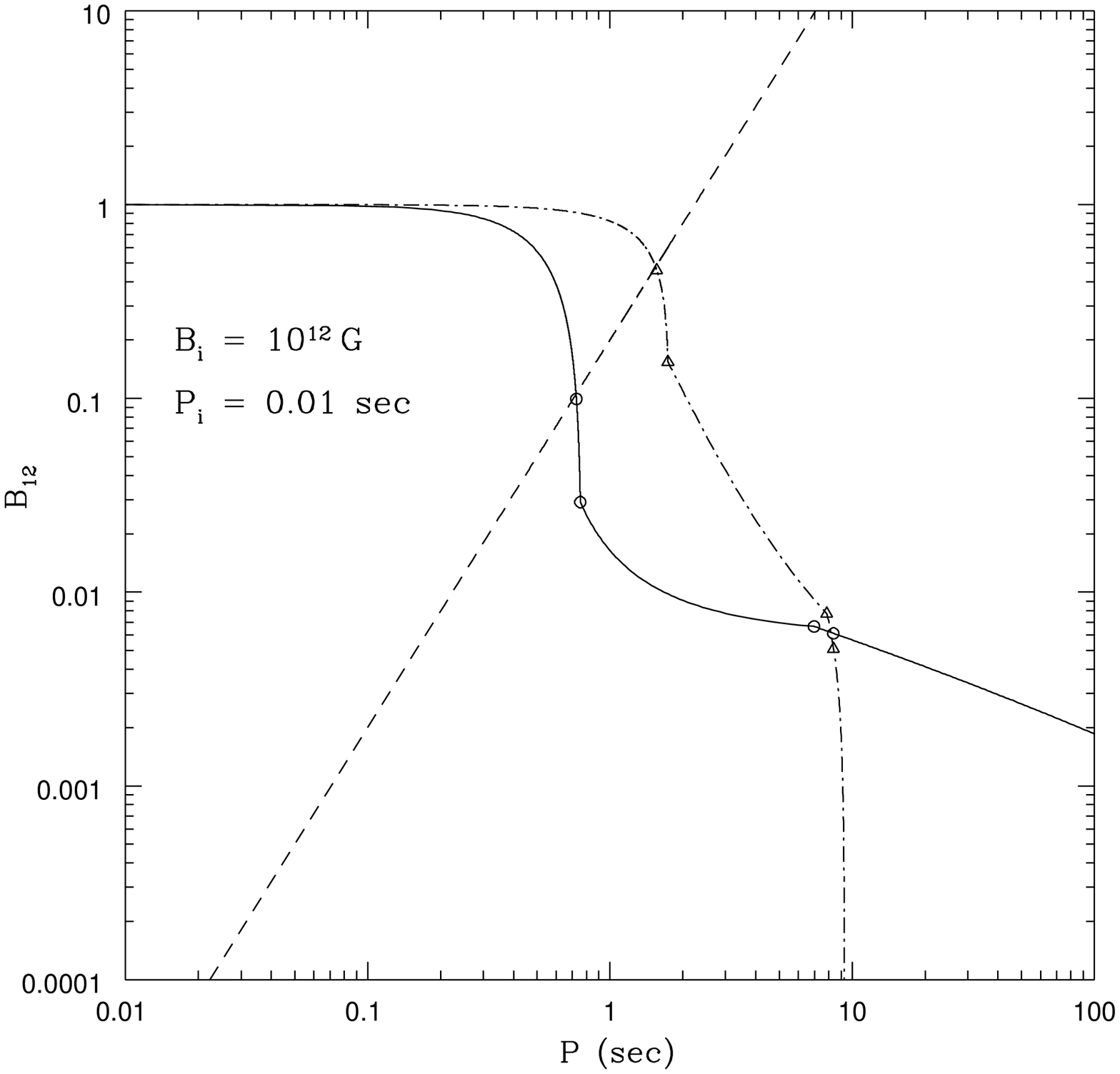}
\end{figure}

\end{document}